# Measurement of carbon finance level and exploration of its influencing factors

Peng Zhang*, Yuwei Zhang, Nuo Xu

*Abstract*：Faced with increasingly severe environmental problems, carbon trading markets and related financial activities aiming at limiting carbon dioxide emissions are booming. Considering the complexity and urgency of carbon market, it is necessary to construct an effective evaluation index system. This paper selected carbon finance index as a composite indicator. Taking Beijing, Shanghai, and Guangdong as examples, we adopted the classic method of multiple criteria decision analysis (MCDA) to analyze the composite indicator. Potential impact factors were screened extensively and calculated through normalization, weighting by coefficient of variation and different aggregation methods. Under the measurement of Shannon-Spearman Measure, the method with the least loss of information was used to obtain the carbon finance index (CFI) of the pilot areas. Through panel model analysis, we found that company size, the number of patents per 10,000 people and the proportion of new energy generation were the factors with significant influence. Based on the research, corresponding suggestions were put forward for different market entities. Hopefully, this research will contribute to the steady development of the national carbon market.

*Keywords*：multiple criteria decision analysis，composite indicator, Shannon-spearman measure，panel regression model

## I. INTRODUCTION

NASA observations from satellite altimeters indicate that the rate of change in global mean sea level is rising anomalously. Together with the frequency of extreme weather, these phenomena could be attributed to global warming caused by high carbon dioxide emissions [1]. Therefore, it is critical to control carbon emissions.

To prevent the climate problem from worsening, the Kyoto Protocol was adopted by 154 countries at the United Nations General Assembly in 1997. Together with the European Union Emissions Trading System (EU ETS) in 2005, the construction of an international carbon trading system was promoted. In recent years, China has taken the construction of a national carbon emissions trading market (hereinafter referred to as the carbon market) as an important initiative to actively respond to climate change and promote the construction of ecological civilization. Since 2011, China has carried out local carbon trading pilot projects in seven areas. Consequently, the total amount and intensity of carbon emissions, covered by the carbon market, have achieved a double reduction. However, the construction of the domestic carbon market is still in its infancy, with insufficient research related to carbon finance. In the face of the complexity of the carbon trading market and a lack of methods to assess the level of carbon finance development, it is necessary to construct a comprehensive evaluation index system.

Recent works mainly focus on four main areas of research related to carbon finance: carbon management, the impact of policies on carbon emissions, the carbon trading market, and the search for optimization strategies through the construction of models. Carbon trading was also studied. Specifically, Wu classified the emission reduction characteristics of each province [2]. Cui et al. studied the current state of carbon trading in the Beijing-Tianjin-Hebei region [3], and Wu et al. simulated the dynamic marginal abatement costs of carbon trading [4]. However, many studies on carbon trading are divided by geography and there is little overall measurement of the level of carbon finance.

Thus, in this paper, we will take carbon finance level as a composite indicator and use classic multiple criteria decision analysis (MCDA) to construct it. The information loss is adopted to measure the weight of different combination methods, while the best aggregation method is chosen to attain the final carbon finance index (CFI), which is firstly adopted together with coefficient of variation method. Furthermore, the influence factors will be screened and analyzed from multiple angles, whose data is obtained through official website of state agencies or commercial databases. Based on the research, this paper will provide corresponding suggestions for power grid companies, power generation enterprises and government. It is believed that measuring and evaluating the level of carbon finance can not only help the construction of a national carbon market system, but also provide a reference basis and an optimization plan for relevant parties when making decisions on their interests.

## II. DEVELOPING CFI BASED ON MULTIPLE CRITERIA DECISION ANALYSIS

Based on the complexity of CFI, MCDA is adopted to construct this composite indicator. Under the Shannon-Spearman Measure-based evaluation model, the performance of MCDA on this issue is scrutinized, followed by choosing the optimal aggregation method, and calculating CFI of all pilot areas.

### A. Composite indicator and multiple criteria decision analysis

A composite indicator (CI) is an index derived from some specific individual indicators for measuring the aggregated performance of a multi-dimensional issue. Technically, the multi-dimensional concepts are measured by a mathematical aggregation of a set of certain indicators, which usually have no common units of measurement [5]. As aforementioned, level of carbon finance is affected by multitudinous factors. Therefore, it is appropriate to regard the CFI as a composite indicator.

*Corresponding@author.email: zhangp_719@outlook.com

The effectiveness of a composite indicator depends heavily on the underlying data aggregation scheme where multiple criteria decision analysis is commonly used. MCDA is a well-established methodology that could instruct policy makers in decision making [6][7]. There are usually three main procedures in construction of a model of m entities and *n* sub-indicators. Firstly, construct the original data matrix *I* and normalize it. Secondly, choose appropriate method to assign weights to different sub-indicators. Thirdly, choose a proper method to complete the aggregation.

### B. Assessment of the composite indicator based on Shannon-Spearman

In CI construction, different MCDA methods affect the overall ramification. While there are many alternative MCDA methods, none could be regarded as a 'super method' suitable for all cases. Based on the framework of Zhou etc., a novel criterion ''information loss'' and an objective measure called the Shannon-Spearman measure (SSM) for comparing MCDA aggregation methods are established [8]. As Zhou et al. argued [9], the SSM is derived based on the loss of information from *I* to CI. The concrete formula is as follows

$$d = \left| \sum_{j=1}^{n} w_j \left(1 + \frac{1}{\ln m} \sum_{i=1}^{m} p_{ij} \ln p_{ij}\right) r_{sj} - \left(1 + \frac{1}{\ln m} \sum_{i=1}^{m} p_i \ln p_i\right) r_s \right| \quad (1)$$

where $p_{ij} = \frac{x_{ij}}{\sum_{i=1}^{m} x_{ij}} (i=1,2,...,m; j=1,2,...,n)$ and $p_i = \frac{CI_i}{\sum_{i=1}^{m} C_i} (i=1,2,...,m)$

It is the aggregation, rather than the normalization and weighting process of original data, that has a significant impact on the CI. Consequently, a MCDA method with a less loss of information is regarded as a better aggregation method (for a perfect aggregation, it is expected that the loss of information could diminish to zero). Therefore, the aggregation method with minimum 'loss information' is selected to construct the final CI.

### C. Construction of primitive CFI based on neoclassical theory

Considering two basic institutions in the market, i.e., production-oriented enterprises and financial institutions. the multi-angle financial indicators were put forward by Chen [10].

Our work selected three certain pilot site ---Beijing, Shanghai, and Guangdong, considering their high economic development, advanced carbon market development and detailed and solid data availability. The carbon emission trading index (CTI) and carbon emission reduction investment index (CII) of the three pilot areas are calculated respectively.

1)Constructing the evaluation system

Inspired by Chen's work, CFI measurement system are explored based on the two sub-CIs of carbon emission trading and carbon reduction investment. Meanwhile, after consulting the climate-related and potential financial impacts for Low Carbon Index released by the World Bank and Trade Map, combined with the characteristics of the overall operating structure of China's Carbon market, four sub-indicators are selected respectively for the two sub-CIs. Since the economic behavior of the carbon market has multiple subjectivity, diversity, and uncertainty, it is reasonable to build the model from the perspective of numerical values instead of analytical expressions, i.e., sub-indicators in every sub-CI are equal at the beginning. The evaluation index system is shown in the figure below.

TABLE I.
EVALUATION INDEX SYSTEM OF CFI

| Evaluation Objections | Evaluation Factors | Indicator Properties |
|---|---|---|
| Carbon Emission Trading | Trading Volume of Carbon Quotas / Number of Enterprises Controlling Carbon Emission | Positive |
| | Trading Volume of CCER / Number of Enterprises Controlling Carbon Emission | |
| | 1/ Standard Deviation of Daily Price | |
| | 1/ Number of Trading Days | Negative |
| Carbon Reduction Investment | (TIs+TIf+TIb) / Number of Enterprises Controlling Carbon Emission | Positive |
| | Number of financial institutions/ (TIs+TIf+TIb) | |
| | Total regional carbon emissions / Number of Enterprises Controlling Carbon Emission | |
| | Weighted loan rate of RMB | Negative |

*Notes: TIs represents total issuance of stocks related to low carbon economy; TIf represents total issuance of funds related to low carbon economy; TIb represents total issuance of bonds related to low carbon economy.*

Specifically, carbon emissions trading is based on the carbon market trading platform. The angles of market depth and validity are both measured, including the unit control line of the scale of enterprise business reflects the market popularization and participation, and the transaction price of standard deviation and trading day days market characteristics features from the perspective of effectiveness. Furthermore, the carbon reduction investment index is examined from the perspective of economics, in which the scale of low-carbon economy can be quantified according to three major financial instruments, namely the total issuance of stocks, funds and bonds related to low-carbon economy. In addition, the total number of financial institutions and the cost of financial services inevitably affect the intensity of investment and finance. Finally, the annual reduction rate of carbon emission intensity is adopted to measure the effectiveness of financial activities. It is worth noting that to ensure the multi-dimension of the measurement, the number of days of carbon trading and weighted loan rate of RMB are selected as the negative variables, as a result, their values become the original reciprocal.

It is a difficult task to accurately describe the carbon emission intensity of pilot areas under various production activities. Thus, the measurement of carbon emission intensity is derived from the perspective of primary energy consumption and electricity consumption with the help of corresponding carbon emission factors. Considering the geographical difference of our three research objects, we refer to the respective carbon intensity, as shown in the following table, further enhancing the accuracy of our model.

TABLE II.
CHINA'S CO2 EMISSION FACTORS

| Energy | Unit | Emission Factor |
|---|---|---|
| *Panel A: Emission Factors of Coal, Oil and Natural Gas* | | |
| Coal | kgCO$_2$/kg | 1.978 |
| Oil | kgCO$_2$/kg | 3.065 |
| Natural Gas | kgCO$_2$/m$^3$ | 1.809 |
| *Panel B: Emission Factors of Electricity* | | |
| North China Grid | kgCO$_2$/kWh | 0.8843 |
| Northeast China Grid | kgCO$_2$/kWh | 0.7769 |
| East China Grid | kgCO$_2$/kWh | 0.7035 |
| Central China Grid | kgCO$_2$/kWh | 0.5257 |
| Northwest China Grid | kgCO$_2$/kWh | 0.6671 |
| China Southern Power Grid | kgCO$_2$/kWh | 0.5271 |

*Source of Panel B: National Center for Climate Change Strategy and International Cooperation, National Development and Reform Commission of China.*

2) Normalizing data and assign weights

Based on the analysis of the above theoretical model, the data of various evaluation factors is obtained through various channels, such as WIND database, China Regional Financial Operation Report and China Energy Statistical Yearbook etc. The first round of data processing is to normalize. At present, three kinds of normalization are commonly used in theoretical research, i.e., linear normalization (LN) method, vector normalization (VN) and multiple attribute utility theory (MAUT). The specific calculation method is shown in Table III.

TABLE III.
THE IMPLEMENTATION FUNCTIONS FOR THE LN AND VN NORMALIZATION METHODS

| Method | Implementation function |
|---|---|
| LN | $r_{ij} = x_{ij}/\max_{i}\{x_{ij}\}, i=1,2,...,m; j=1,2,...,n$ |
| VN | $r_{ij} = x_{ij}/\sqrt{\sum_{i=1}^{m} x_{ij}^2}, i=1,2,...,m; j=1,2,...,n$ |
| MAUT | $r_{ij} = \dfrac{\max\{x_{ij}\} - x_{ij}}{\max\{x_{ij}\} - \min\{x_{ij}\}}, i=1,2,...,m; j=1,2,...,n$ |

The last one has been widely used in the initial normalization of data because it considers characteristics of all data without deliberately amplifying characteristics of one data point. Intuitively, the corresponding normalization method is adopted. To avoid the frequent occurrence of *0, 1* after normalization, the global maximum and minimum values are selected. The implementation functions for benefit sub-indicator (the lager the better) are shown as follows

$$X_{ijt}^{+} = \frac{x_{ijt} - m_i}{M_i - m_i}$$
$$X_{ijt}^{-} = \frac{M_i - x_{ijt}}{M_i - m_i}$$
(2)

where, *i is the evaluation factor; j is the pilot area;* $X_{ijt}^{+}$ and $X_{ijt}^{-}$ *represent the dimensionless methods for positive and negative indicators, repectively;* $x_{ijt}$ *is the initial value;* $M_{ij}$ and $m_{ij}$ *represent the maximum and minimum value of this indicator in each pilot area.*

Next, weighting operation is carried out for evaluation systems constructed from different perspectives. The *coefficient of variation method* calculated based on statistical methods is adopted. The index with large variation difference has a large weight. The specific weight calculation formula is as follows.

$$\omega_{it} = V_{ijt}/\sum_{i} V_{ijt} \quad V_{ijt} = \sigma_{ijt}/\bar{X}_{ijt}$$ (3)

where, $\omega_{it}$ *represents the weight of evaluation factor i in year t;*

$V_{ijt}$ *represents the coefficient of variation of the index;*

$\sigma_{ijt}$ *represents the standard deviation of the metric after dimensionless;*

$\bar{X}_{ijt}$ *represents the meanvalue of the metric after dimensionless.*

D. *Conducting distinct aggregation methods*

Five alternative MCDA aggregation methods are chosen, namely the simple additive weighting (SAW), the weighted product (WP), the weighted displaced ideal (WDI) with parameters 2 (hereafter called WDI2) and ∞ (hereafter called WDI∞), and the TOPSIS methods, which have been widely studied in CI construction [5][11][12]. It is noted that the TOPSIS method has been barely adopted in construct CI, however it is equipped with some attractive properties [7]. Table IV shows the concrete formula for the five distinctive aggregation methods.

TABLE IV.
THE AGGREGATION FUNCTIONS FOR FIVE ALTERNATIVE MCDA METHODS

| Method | Aggregation function |
|---|---|
| SAW | $\text{CI}_i = \sum_{j=1}^{n} w_j r_{ij}, i=1,2,...,m$ |
| WP | $\text{CI}_i = \prod_{j=1}^{n} (r_{ij})^{w_j}, i=1,2,...,m$ |
| WDI2 | $\text{CI}_i = \sqrt{\sum_{j=1}^{n}(w_j r_{ij})^2}, i=1,2,...,m$ |
| WDI ∞ | $\text{CI}_i = \min_{j}\{w_j r_{ij}\}, i=1,2,...,m$ |
| TOPSIS | $CI_i = \dfrac{\sqrt{\sum_{j=1}^{n}\left(w_j r_{ij} - \min_{i}\{w_j r_{ij}\}\right)^2}}{\sqrt{\sum_{j=1}^{n}\left(w_j r_{ij} - \min_{i}\{w_j r_{ij}\}\right)^2} + \sqrt{\sum_{j=1}^{n}\left(w_j r_{ij} - \max_{i}\{w_j r_{ij}\}\right)^2}}, i=1,2,...,m$ |

According to Table IV, all the numerical calculation are based on MATLAB, the results of the CTI and CII under different methods are shown in Table V and Table VI, respectively.

TABLE V.
CTI CALCULATED FROM FOUR DIFFERENT AGGREGATION METHOD

| 2015 | SAW | WP | WDI2 | WDI∞ | TOPSIS |
|---|---|---|---|---|---|
| BJ | 0.3941 | 0.3305 | 0.2284 | 0.0393 | 0.4940 |
| GD | 0.2591 | 0 | 0.1833 | 0 | 0.4284 |
| SH | 0.4615 | 0.2342 | 0.3766 | 0.006 | 0.3853 |
| …… | | | | | |
| 2020 | | | | | |
| BJ | 0.0994 | 0 | 0.0738 | 0 | 0.4119 |
| GD | 0.5217 | 0.4782 | 0.0295 | 0.0295 | 0.4615 |
| SH | 0.3022 | 0 | 0.2251 | 0 | 0.4995 |

TABLE VI.
CII CALCULATED FROM FOUR DIFFERENT AGGREGATION METHOD

| 2015 | SAW | WP | WDI2 | WDI∞ | TOPSIS |
|---|---|---|---|---|---|
| BJ | 0.4326 | 0.3690 | 0.2482 | 0.0467 | 0.4852 |
| GD | 0.4314 | 0.2321 | 0.2954 | 0.0084 | 0.4546 |
| SH | 0.3930 | 0.2568 | 0.2379 | 0.0099 | 0.5243 |
| …… | | | | | |
| 2020 | | | | | |
| BJ | 0.2910 | 0 | 0.2116 | 0 | 0.4168 |
| GD | 0.4263 | 0.3581 | 0.2688 | 0.0543 | 0.3725 |
| SH | 0.2009 | 0.0560 | 0.1196 | 0.0013 | 0.5326 |

### E. Evaluating performance

Based on the above results, information loss is used to quantify the corresponding performances. The first part is measured by Shannon entropy, and the probability of the sub-indicator is shown in Eq. (4). The second part is derived from the ranking of $n$ sub-indicators and the CI Derived. The Spearman rank correlation coefficient $RSJ$ and $RS$ between them and a reference rank sequence such as $r_0 = (m, m-1, ...)$ shown in Eq. (5). In the third part, the weight matrix is extracted based on the weight assigned and combined with the variation coefficient method we have chosen.

$$p_{ij} = \frac{x_{ij}}{\sum_{i=1}^{m} x_{ij}} (i=1,2,...,m; j=1,2,...,n) \text{ and } p_i = \frac{CI_i}{\sum_{i=1}^{m} C_i} (i=1,2,...,m) \quad (4)$$

where, $d_i$ represents the difference between the two ranks of each observation; $n$ is the number of observations.

Technically, it denotes the usual Pearson correlation coefficient of the rank variables. $cov(rg_X, rg_Y)$ is the covariance of the rank variables, while $\sigma_{rgX}$ and $\sigma_{rgY}$ are the standard deviations of the rank. Finally, $r_s$ is computed in Eq. (6).

$$r_s = \rho_{rgX, rgY} = \frac{cov(rg_X, rg_Y)}{\sigma_{rgX} \sigma_{rgY}} \quad (5)$$

$$r_s = 1 - \frac{6 \sum d_i^2}{n(n^2 - 1)} \quad (6)$$

Considering its characteristics of time series, the losses of the five aggregation methods in each year during 2015-2020 are calculated, the obtained results are shown in Table VII by taking the arithmetic average. It is transparent that WDI2 has the minimum information loss function value, therefore, this method is chosen as the final aggregation method of our model.

TABLE VII.
MEAN OF D OF THREE PILOT AREAS OVER 2015-2020

| | SAW | WP | WDI2 | WDI∞ | TOPSIS |
|---|---|---|---|---|---|
| dmean | 0.0352 | 0.0287 | 0.0254 | 0.0278 | 0.0266 |

### F. Calculate final CFI

Based on the WDI2 method, the CTI index for the three pilot regions is calculated, as shown in Table VIII and Fig.1.

TABLE VIII.
CTI FOR THE THREE PILOT REGIONS FROM 2015 TO 2020

| Region | 2015 | 2016 | 2017 | 2018 | 2019 | 2020 | Regional average |
|---|---|---|---|---|---|---|---|
| BJ | 0.228 | 0.168 | 0.177 | 0.220 | 0.237 | 0.074 | 0.184 |
| GD | 0.183 | 0.244 | 0.247 | 0.224 | 0.269 | 0.327 | 0.249 |
| SH | 0.377 | 0.202 | 0.220 | 0.208 | 0.225 | 0.181 | 0.235 |
| Period average | 0.263 | 0.205 | 0.215 | 0.217 | 0.244 | 0.194 | 0.223 |
| Standard deviation | 0.101 | 0.038 | 0.035 | 0.008 | 0.023 | 0.127 | |

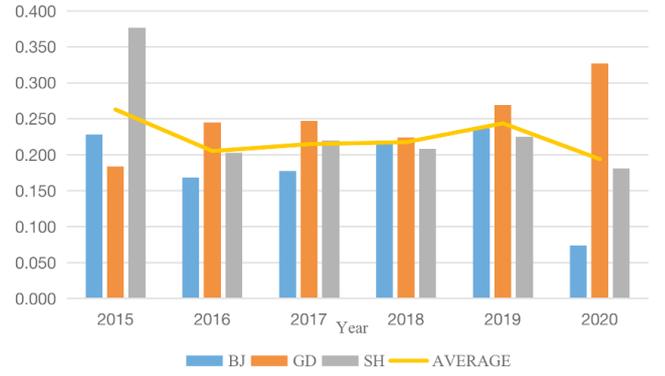

Fig.1. CTI of three pilot areas

From Table VIII and Fig.1, the average CTI values for the three pilots, except in 2015 when they exceeded 0.25 and in 2020 when they were below 0.2, remained stable between 0.2 and 0.25 in all four years, indicating that overall CTI development was relatively unruffled. The remaining four years are all stable in the range of 0.2-0.25, indicating that overall CTI of the three pilot is developing more slowly. In terms of period averages, Guangdong has the highest level of carbon emissions trading development, with the average CTI value reaching 0.209. With the stabilization above 0.2 after 2018, the trend of steady growth is more obvious.

Results of the CII index are shown in Table IX and Fig.2.

TABLE IX.
CII FOR THE THREE PILOT REGIONS FROM 2015 TO 2020

| Region | 2015 | 2016 | 2017 | 2018 | 2019 | 2020 | Regional average |
|---|---|---|---|---|---|---|---|
| BJ | 0.248 | 0.182 | 0.269 | 0.293 | 0.144 | 0.212 | 0.225 |
| GD | 0.295 | 0.349 | 0.283 | 0.233 | 0.287 | 0.269 | 0.286 |
| SH | 0.238 | 0.219 | 0.193 | 0.188 | 0.188 | 0.120 | 0.191 |
| Period average | 0.261 | 0.250 | 0.249 | 0.238 | 0.206 | 0.200 | 0.234 |
| Standard deviation | 0.031 | 0.088 | 0.048 | 0.052 | 0.073 | 0.075 | |

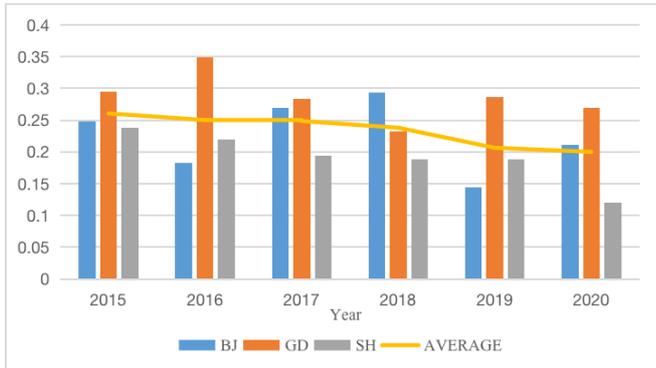
Fig.2. CII of three pilot areas

As seen in the Table IX and Fig.2, in terms of period averages, 2015 was the maximum among the three pilot regions, reaching 0.261, and in 2020 it reached 0.2, the minimum value. It indicates that the annual CII of the three regions has a downward trend but is relatively stable. Among the three pilots, Guangdong's advantage is still more obvious, with its CII being the highest among the three pilots in all years except 2018, and the regional average value is also the highest among the three pilot regions. The annual CII value in Beijing is more volatile, while the annual CII value in Shanghai has a slowly declining trend.

Further, we repeat the assignment and aggregation operation from evaluation factors to evaluation targets (CTI & CII), and finally obtain the CFI for Beijing, Shanghai, and Guangdong, as shown in Table X and Fig.3.

TABLE X.
CFI FOR THE THREE PILOT REGIONS FROM 2015 TO 2020

| Region | 2015 | 2016 | 2017 | 2018 | 2019 | 2020 | Regional average |
|---|---|---|---|---|---|---|---|
| BJ | 0.184 | 0.133 | 0.167 | 0.251 | 0.124 | 0.090 | 0.158 |
| GD | 0.156 | 0.243 | 0.191 | 0.201 | 0.234 | 0.230 | 0.209 |
| SH | 0.294 | 0.159 | 0.145 | 0.163 | 0.156 | 0.123 | 0.173 |
| Period average | 0.211 | 0.178 | 0.168 | 0.205 | 0.172 | 0.148 | 0.180 |
| Standard deviation | 0.073 | 0.058 | 0.023 | 0.044 | 0.056 | 0.073 | |

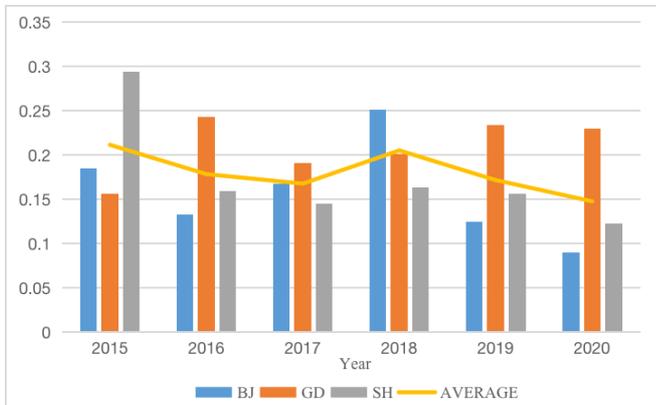
Fig.3. CFI of three pilot areas

As seen in the Table X and Fig.3, in terms of period averages, the maximum value was reached at 0.211 in 2015, and exceeded 0.2 in both 2015 and 2018, with an overall stable development trend despite ups and downs. Regarding the regional averages, Guangdong is still the region with the highest level of carbon finance development among the three pilot regions.

III. ANALYSIS OF INFLUENCING FACTORS BASED ON PANEL REGRESSION MODEL

After obtaining the CFI of three pilot areas, we further constructs a panel regression model to empirically analyze the influencing factors of CFI. The factors selected in this paper include the following four aspects, the proportion of added value of the secondary industry, the number of patents per 10000 people, the scale of enterprises, the geographical location of the pilot and the proportion of new energy power generation.

*A. Variable selection*

1) CFI.

The statistics calculated in the previous part are used as the measurement of CFI.

2) Proportion of secondary industry's added value

The index of "the added value of the secondary industry/ the added value of GDP" is selected to measure the proportion of secondary industry's added value. Since most of the secondary industry belongs to heavy industry, the increase of industrial added value caused by the development of heavy industry will increase the degree of pollution and carbon dioxide emission.

3) The Number of patents per 10000 people

The more patents per 10000 people have, the stronger the scientific research ability of the pilot will be equipped with. Consequently, the R&D capability of low carbon technology will be enhanced and the inventions of low carbon products will increase, resulting in the augment of the investment and financing activities of financial institutions.

4) Company size

The operating income of industrial enterprises above designated size is used as the measurement standard in this paper. Large enterprises are more likely to have stronger comprehensive strength, which can exert a positive impact on carbon emissions in the production process of enterprises, as well as carbon finance related investment and financing activities.

5) Location

The geographical location of the city is crucial in its own development. With convenient transportation, developed foreign trade and other resources, coastal areas develop more rapidly compared with inland areas. Geographical location is chosen as a virtual variable to explore the impact of coastal or inland on the development of carbon finance. "1" represents coastal areas and "0" represents inland areas.

6) Proportion of new energy generation

New energy power generation is generating electricity through using new energy such as solar energy, wind energy etc. The percentage of power generation except thermal power generation in total power generation is chosen to represent the proportion of new energy generation.

A summary of these variables is depicted in the Table XI.
TABLE XI.
DEFINITIONS OF VARIABLES

| Variable name | Variable symbol | Variable description |
|---|---|---|
| Carbon finance index | CFI | The statistics calculated above |
| Proportion of secondary industry's added value | PSI | The added value of the secondary industry/ the added value of GDP |
| The number of patents per 10000 people | Patent | Relevant data in statistical yearbooks of cities |
| Company size | Size | Operating income of industrial enterprises above scale |
| Location | Location | Virtual variable "1" represents coastal areas and "0" represents inland areas |
| Proportion of new energy generation | Energy | Other power generation (except thermal power generation) / total generation |

*B. Regression model setting*

To analyze the influencing factors of carbon finance, the following panel regression model is established.

$$CFI_{it} = \alpha_0 + \alpha_1 IVA_{it} + \alpha_2 Patent_{it} + \alpha_3 Size_{it} + \alpha_4 Location_{it} + f_i + d_t + \varepsilon_{it} \quad (7)$$

where $i$ and $t$ represent the pilot and time, respectively.

*C. Empirical Analysis*

1) Descriptive statistics

The descriptive statistics of each variable are shown in Table XII. From 2014 to 2020, the minimum value of CFI is 0.154187, and the maximum value is 0.711969. Thus, there are great differences in the degree of carbon finance development among these regions. The average value of the proportion of the secondary industry's added value is 0.321543, which reflects that the increase of GDP in several regions mainly depends on the secondary industry. The average number of patents per 10000 people is 9.707222, and the median is 9.730000, meaning that the level of scientific and technological innovation in each pilot is not significantly different. The minimum value of company size is 19256.10, while the maximum value is 149930.1, denoting that there are large differences in the size of firms and carbon emissions. The average value of the proportion of new energy generation is 0.120437, indicating that thermal power generation is especially dominant.

TABLE XII.
DESCRIPTIVE STATISTICS OF MAIN VARIABLES

| Variables | Mean value | Standard deviation | Minimum value | Median | Maximum value |
|---|---|---|---|---|---|
| CFI | 0.180280 | 0.053018 | 0.090296 | 0.165106 | 0.293644 |
| PSI | 0.321543 | 0.192219 | 0.158338 | 0.287230 | 0.973230 |
| Patent | 9.707222 | 3.592854 | 3.980000 | 9.730000 | 16.30000 |
| Size | 65422.84 | 52858.88 | 19256.10 | 38398.46 | 149930.1 |
| Energy | 0.120437 | 0.131243 | 0.008857 | 0.042052 | 0.359048 |

2) Analysis of empirical regression results

To determine the model selection, the Hausman test is carried out on the sample data to examine the adaptability of the fixed effect model and the random effect model. The test results are shown in Table XIII.

TABLE XIII.
HAUSMAN TEST

| Test Summary | Chi-Sq. Statistic | Chi-Sq. d.f. | Prob. |
|---|---|---|---|
| Period random | 0.000000 | 5 | 1.0000 |

It can be seen that the adjoint probability is 1.0000. Therefore, a random effect model is established. The regression results are shown in the table below.

TABLE XIV.
REGRESSION RESULTS

| Variable | CFI |
|---|---|
| PSI | -0.114241 |
|  | (-0.503886) |
| Patent | 0.010701** |
|  | (2.222803) |
| Size | 0.0000026*** |
|  | (3.107416) |
| Location | 0.024281 |
|  | (0.491994) |
| Energy | 0.775972*** |
|  | (4.489495) |
| C | 0.367741*** |
|  | (3.028600) |
| R-squared | 0.751765 |
| Adjusted R-squared | 0.675385 |
| F-statistic | 9.842412 |
| Prob(F-statistic) | 0.000686 |

*Notes: *, * *, * * * represent the regression coefficient is significant at the level of 10%, 5% and 1% respectively; () represents the T value of each regression coefficient. The same below.*

The results show that from the overall perspective of the equation, the probability P value is 0.000686, which is significant at the level of 1%. The adjusted goodness of fit Adjusted R-squared is 0.675385. For the panel data, the overall fitting effect is very good. In summary, the model constructed in this paper is reasonable, as well as the selected data samples. However, only the coefficients of enterprise size and the proportion of new energy generation are significant respectively at the level of 1%, with the coefficient of the number of patents per ten thousand people significant at the level of 5%.

The possible explanations for these phenomena are as follows. Since the "operating income of industrial enterprises above scale" is used as the measurement of enterprise scale, the operating income also reflects their profitability. Enterprises with great operating income and high profitability can use redundant funds for R&D to explore production and operation modes after maintaining their normal operation, which is conducive to reducing carbon emissions, or actively participating in investment and financing activities related to carbon finance. All the measures above can promote the overall level of carbon finance in the pilot areas. On the other hand, the "number of patents per 10,000 people" reflects the scientific and technological innovation capability of the region. The more patents, the higher the R&D capability of the region. Consequently, the exploration of low carbon technology is propelled, resulting in the increase of low carbon products. Therefore, the investment and financing activities of financial institutions in the region will be improved accordingly, which promotes the CFI of the pilot. The growing proportion of new energy generation indicates that carbon emissions have been reduced, representing a high efficiency in the use of carbon funds and a fabulous development of carbon finance. In summary, the proportion of new energy generation promotes the development of carbon finance.

## IV. DISCUSSION AND SUGGESTIONS

Through the analysis of the carbon finance index obtained by building a panel model, it is found that the main influencing factors are *company scale, the number of patents per 10,000 people, and the proportion of new energy generation*. Further, it addresses the possible explanations for these phenomena.

Firstly, the operating income not only reflects the scale of enterprise, but also their profitability. For those enterprises with great operating income and high profitability, they can use redundant funds for R & D to explore production and operation modes which are conducive to reducing carbon emissions. Therefore, the scale of enterprises plays a positive role in promoting the level of carbon finance. It is consistent with the national level to limit carbon emissions, develop green finance and green electricity policies to take the lead from power generation enterprises. Thus, it highlights the leading role of large state-owned enterprises in promoting the development of carbon finance and reducing environmental pressure.

Secondly, "the number of patents per 10,000 people" reflects the scientific and technological innovation capability of the region. At the same time, the research ability of low carbon technology is also enhanced, resulting in the increase of low carbon products. Therefore, the investment and financing activities of financial institutions in the region will be improved accordingly, which promotes the CFI of the pilot. This influencing factor is closely related to China's construction of a scientific and technological power and technology-oriented healthy economy by developing key scientific research in key areas to catch up with and even surpass the international advanced level.

Lastly, the incremental proportion of new energy generation reduces the carbon emissions, which demonstrates the use efficiency of carbon funds and the effective development of carbon finance. This is a perfect example of our current efforts to develop new energy sources and increase their share of power generation. In the next step, relevant departments should continue to increase investment in new energy, overcome multiple difficulties such as grid connection, absorption, and instability of new energy, constructing a solid foundation for the carbon peak in 2030 and carbon neutralization in 2060.

## V. CONCLUSION

In this paper, the carbon finance level evaluation system was constructed through MCDA. Eight sub-indicators including *Trading Volume of Carbon Quotas*, *Total Regional Carbon Emissions* etc. were selected. Considering different normalized characteristics, the classic MAUT normalized method was primarily chosen. Furthermore, the coefficient of variation method was used to assign weights to each factor. Then, five different aggregation methods were compared, while the optimal combination method was determined based on Shannon-Spearman Measure information loss. At last, a complete carbon finance index was obtained.

A panel model with proportion of secondary industry's added value production floor and company size etc. as variables was constructed. Through the analysis and calculation, three significant influencing factors were obtained, *Company scale, the number of patents per 10,000 people, and the proportion of new energy generation*. Furthermore, we put forward relevant opinions for relevant departments and enterprises, which will contribute to the steady operation and level evaluation of national carbon market.

It is noted that improvement and generalization could be conducted better under the sufficient data among all the pilot areas. Considering that the establishment of the national carbon market has already started, it is promising to extend our model to a large-scale carbon financial evaluation system.

## APPENDIX

*Sources of data:*

*Trading volume of carbon emission quotas, standard deviation of daily and number of trading days* are from the CSMAR. *Trading volume of CCER* is from the pilot regional exchanges. *Total number of enterprises controlling carbon emissions* is from the data published by Local Development and Reform Commission.

For "*Stocks*", we select "Energy Saving and Environmental Protection", "Green Cleaning", "Environmental Protection Concept", "New Energy Vehicles" and other related sectors from the WIND database. For "*Funds*", we select "Wind Power", "Green Energy Saving", "Environmental Protection Concept", "Low Carbon Environmental Protection" and other related sectors from the WIND database. For "*Bonds*", select "Carbon Neutral", "Green Finance", "Energy Saving and Environmental Protection" and other related sectors from the WIND database.

Data on the *annual carbon intensity* are obtained from the China Energy Statistic Yearbook. Data on *weighted loan rate of RMB* and total number of financial institutions are acquired from the Regional Financial Performance Report.

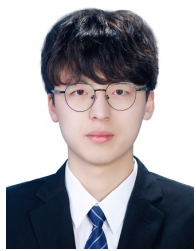
Peng Zhang* was born in Zhangjiakou, China in 2000. He is an undergraduate student, majoring in Electrical Engineering and Its Automation, in School of Electrical and Electronic Engineering, North China Electric Power University, Beijing, China.
(e-mail: zhangp_719@outlook.com)

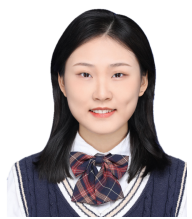
Yuwei Zhang was born in Xuzhou, China, in 2000. She is an undergraduate student, majoring in Finance, in School of Economics and Management, North China Electric Power University, Beijing, China.
(e-mail: zyw9992977@126.com)

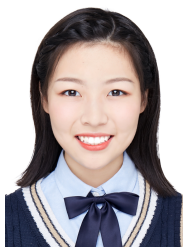
Nuo Xu was born in Linfen, China, in 2000. She is an undergraduate student, majoring in Law, in School of Humanities and Social Sciences, North China Electric Power University, Beijing, China.
(e-mail: promise010622@163.com)